\documentclass{article}
\usepackage{spconf,amsmath,graphicx}
\usepackage{multirow,booktabs}
\usepackage{makecell}
\title{ Robust Multi-channel Speech Recognition \\ using Frequency Aligned Network}

\name{Taejin Park$^{1*}$\thanks{*Taejin Park completed this work while interning at Amazon.}
, Kenichi Kumatani$^{2}$, Minhua Wu$^{2}$, Shiva Sundaram$^{2}$} 

\address{$^1$ University of Southern California (USC), Los Angeles, CA, USA\\
$^2$Amazon Inc., Sunnyvale, CA, USA\\}
\begin{document}
\ninept

%
\maketitle
\begin{abstract}
Conventional speech enhancement technique such as beamforming has known benefits for far-field speech recognition. Our own work in frequency-domain multi-channel acoustic modeling has shown additional improvements by training a spatial filtering layer jointly within an acoustic model. In this paper, we further develop this idea and use frequency aligned network for robust multi-channel automatic speech recognition (ASR). Unlike an affine layer in the frequency domain, the proposed frequency aligned component prevents one frequency bin influencing other frequency bins. We show that this modification not only reduces the number of parameters in the model but also significantly and improves the ASR performance. We investigate effects of frequency aligned  network through ASR experiments on the real-world far-field data where users are interacting with an ASR system in uncontrolled acoustic environments. We show that our multi-channel acoustic model with a frequency aligned network shows up to 18\%  relative reduction in word error rate.
\end{abstract}
\section{Introduction}
\label{sec:intro}

 
Multi-channel ASR has been considered as an effective way to cope with noise and reverberation for a distant speech recognition (DSR) system. In general, voice activity detection, dereverberation and beamforming play an integral part of a DSR system that processes far-field speech. Beamforming techniques, take advantage of multiple microphones to enhance the audio signal. Beamforming can be categorized into fixed beamforming or adaptive beamforming. In comparison to fixed beamforming, adaptive techniques have shown that noise robustness of ASR system can be improved with a dereverberation  approach \cite{delcroix2014linear} or high-order statistics \cite{kumatani2012microphone}. However, adaptive techniques rely on accurate voice activity detection or speaker location estimation, and therefore they  can underperform in comparison to a fixed beamformer; especially when these dependent components are not performing reliably \cite{sullivan1996multi, himawan2010clustered}. Further, according to the previous studies by McDonough et al. \cite{mcdonough2008distant} and Seltzer et al. \cite{seltzer2008bridging}, individually optimizing various DSR components turns out to be a sub-optimal solution.

More recently, multi-channel deep neural network (MC-DNN) approaches have been applied to a multi-channel ASR by training a unified MC-DNN  model where  the multi-channel processing modules are part of the DNN structure. The work by Xiao et al. \cite{xiao2016deep} proposed a unified system that incorporates  beamforming, feature extraction, and acoustic modeling into a unified model. The parameters of a beamformer is first estimated and then applied to the array signals to enhance the signal. Thus, the system in \cite{xiao2016deep} is not a fully learnable system. The end-to-end system proposed by Ochiai et al. \cite{ochiai2017multichannel} also contains jointly optimized beamforming components. However, this end-to-end system requires bidirectional LSTM that makes the system inapplicable to real-time system. The methods presented in \cite{swietojanski2014convolutional, braun2018multi, kim2015recurrent} are also unified MC-DNN ASR systems but they are not fully learnable systems. Aside from unified MC-DNN approaches, a DNN is also employed to construct a clean speech signal. An LSTM mask based method was proposed to estimate the statistics of target \cite{heymann2018performance, higuchi2018frame}. However, mask based beamforming approach needs accumulated statistics from adequate amount of adaptation data to show the improvement. Accumulating the statistics might cause additional latency to the system by requiring more memory usage and often not applicable to real-time applications.

In \cite{minhua2019frequency}, we proposed a fully learnable multi-channel acoustic model where array processing front-end was integrated into our trainable neural network module. The proposed acoustic model improved from single channel model showing that fully learnable multi-channel ASR system benefits the speech recognition performance in terms of word error rate (WER). In addition, the acoustic model supports real-time application and the DNN architecture does not involve any module that requires accumulation of statistics or  bidirectional processing of a speech utterance.
While our multi-channel model shows significant improved performance, we still see reduced improvement in  the presensce of intense background noise. The most prominent example for such a condition is the presence of playback echo. Playback echo is a self-generated acoustic signal from a device where the acoustic signal is also picked up by the microphones on the device. Playback echo is one of the most challenging signal processing problem since the signal has relatively high amplitude when it is generated in close proximity to the microphones.

In this work, to improve the noise robustness and the overall ASR accuracy, we propose frequency aligned network (FAN) architecture that independently processes each frequency bin and enables frequency independent processing that confines the influence of frequency bin input to each frequency bin. The FAN architecture is built over our previous multi-channel ASR framework \cite{minhua2019frequency} and it has a couple of practical benefits. First, FAN  requires far fewer parameters, compared to a vanilla affine layer and thus lead to lower complexity of calculation and less memory requirements. Second, the average pooling layer of FAN gives a good performance gains for utterances degraded by playback echo. These gains were observed in contrast to max pooling approach where we implicitly select only one look direction.

We show the benefit of the proposed FAN architecture through ASR experiments on an in-house multi-channel speech dataset that also included interference due to playback echo and noise. The dataset includes speech utterances acquired in unconstrained acoustic environments where volunteers are interacting with our Alexa devices on a daily basis. 
\begin{figure}[t]
  \centering
  \centerline{\includegraphics[width=8.0cm]{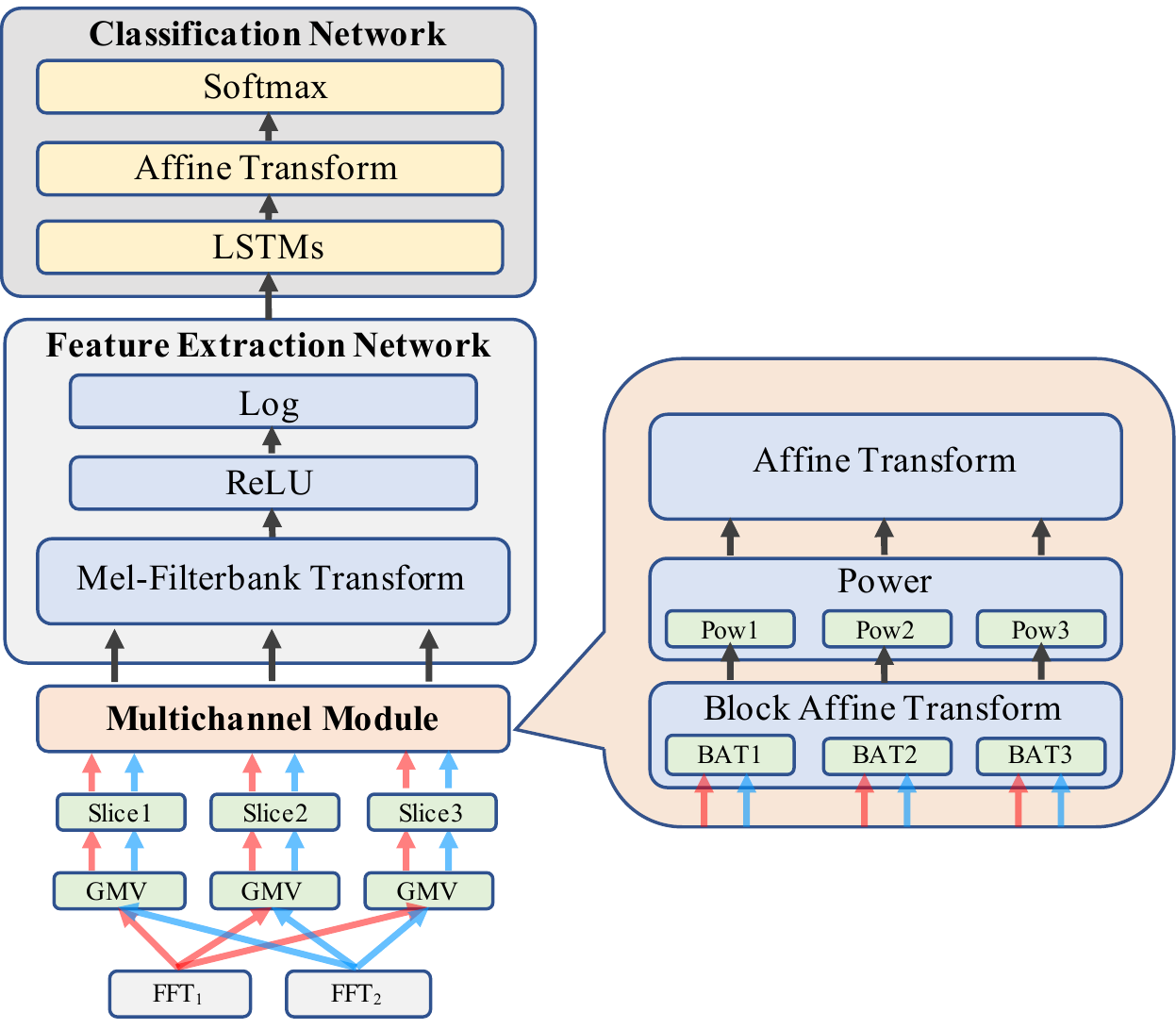}}
    \centerline{}
    \vspace{-5.0ex}
\caption{multi-channel ASR framework and the multi-channel modules}
    \vspace{-3.0ex}
\label{fig:baseline}
\end{figure}

\section{Baseline Multichannel DSR System}
In \cite{minhua2019frequency}, we introduced a two-channel system with a fully learnable front-end.  We showed that our two-channel acoustic model yields over 9\% WER reduction over conventional beamforming front-end using seven microphones. In this work, we employ a similar acoustic model but a low frame rate (LFR) \cite{pundak2016lower} version.  Through LFR, we lower the scoring rate from 10 msec to 30 msec by feeding three consecutive frames simultaneously. All the components described in this paper are therefore based on LFR approach where every component has three separate signal stream that enables 1/3$^{rd}$ rate of a regular frame rate system. Figure \ref{fig:baseline} illustrates this model.

The acoustic model consists of 4 different stages: pre-processing, multi-channel (MC) layer, feature extraction (FE) and stacked LSTMs for senone classification.  First, we extract discrete Fourier Transform features (DFT) from the audio frames from each channel using Fast Fourier Transform (FFT). We then apply global mean and variance normalization (GMVN) in the frequency domain to each channel before passing it to the MC layer. For the MC layer, we used a block-wise affine transform  combined with a vanilla affine layer in \cite{minhua2019frequency} (elastic spatial filtering).  As discussed later, in this work, we evaluate different strategies for this block. The output of MC module is connected to FE layer. The  FE-DNN layer is designed to output log-filter bank energy (LFBE) feature. First, the input signal goes into an affine layer initialized with Mel filterbank weights.  The output of the filterbank goes into Rectified linear unit (ReLU) and log function. The ReLU component is employed to enforce the positive number for the log function. The FE-DNN is initialized to mimics (LFBE) feature, but like other components, it is fully trainable. The output from FE-DNN  goes to the classification network.
We train the weights of the network in a stage-wise manner \cite{kumatani2017direct}. We first built the classification LSTM with
the single channel LFBE feature, then trained the cascade network of the FE and classification layers with the single-channel DFT feature, and finally performed joint optimization on the whole network with multi-channel DFT input. In contrast to the conventional DSR system, this fully learnable acoustic model approach neither requires clean speech signal reconstruction nor perceptually-motivated filter banks \cite{richard2013perceptual}.

In this paper, we focus on the performnace of MC module in Fig \ref{fig:baseline}. We compare the performance of our proposed FAN architecture with linear layer with single-channel beamformed input. We focus on the block affine transform (BAT) that showed superior performance in our previous study \cite{minhua2019frequency}. The variants of the MC components studied in this paper are discussed next.  
\section{Frequency Aligned Network in Multi-channel Module}
\label{sec:format}

\subsection{Block Affine Transform}
In our previous study, we proposed the fully learnable MC modeling method that gives an improvement over a beamforming solution \cite{minhua2019frequency}. This \textit{elastic spatial filtering} (SF) includes the block affine transforms initialized with super directive beamformers’ weights to form 12 different look directions. In doing so, we incorporate array processing knowledge into fully learnable MC module. Especially in the elastic SF module, the 12 look directions can be learned through the training process.  In this work, we use BAT module and power module which are parts of elastic SF module. The discrete fourier transform of input signal can be described as below:
\begin{equation}
 \mathbf{X}\left(t, \omega_{k}\right)=\left[X_{1}\left(t, \omega_{k}\right), \cdots, X_{M}\left(t, \omega_{k}\right)\right]^{T}
\end{equation} 
Using this notation, we can express complex weight vector for source position $\mathbf{p}$ as follows:
\begin{equation}
\mathbf{w}\left(t, \omega_{k}, \mathbf{p}\right)=\left[w_{1}\left(t, \omega_{k}, \mathbf{p}\right), \cdots, w_{M}\left(t, \omega_{k}, \mathbf{p}\right)\right]
\end{equation}
Thus, BAT and the following power operation can be expressed as follows:
\begin{equation}\footnotesize
\label{eq:bat}
\left[\begin{array}{c}{Y_{1}\left(\omega_{1}\right)} \\ {\dots} \\ {Y_{D}\left(\omega_{1}\right)} \\ {\dots} \\ {Y_{1}\left(\omega_{K}\right)} \\ {\dots} \\ {Y_{D}\left(\omega_{K}\right)}\end{array}\right] = \text{pow} \left(
\left[\begin{array}{c}{\mathbf{w}_{\mathrm{SD}}^{H}\left(\omega_{1}, \mathbf{p}_{1}\right) \mathbf{X}\left(\omega_{1}\right)+\mathbf{b}_{1}} \\ {\dots} \\ {\mathbf{w}_{\mathrm{SD}}^{H}\left(\omega_{1}, \mathbf{p}_{D}\right) \mathbf{X}\left(\omega_{1}\right)+\mathbf{b}_{D}} \\ {\dots} \\ {\mathbf{w}_{\mathrm{SD}}^{H}\left(\omega_{K}, \mathbf{p}_{1}\right) \mathbf{X}\left(\omega_{K}\right)+\mathbf{b}_{D K}} \\ {\dots} \\ {\mathbf{w}_{\mathrm{SD}}^{H}\left(\omega_{K}, \mathbf{p}_{D}\right) \mathbf{X}\left(\omega_{K}\right)+\mathbf{b}_{D(K+1)}}\end{array}\right] \right)
 \normalsize
\end{equation} where  $\mathbf{b}$ is bias term, $D$ is the number of look directions and $K$ is the number of frequency bins. After power operation in eq. \ref{eq:bat}, 254 dimensional DFT feature is converted to 127 dimensional vector.

\begin{figure}[t]
  \centering
  \centerline{\includegraphics[width=8.2cm]{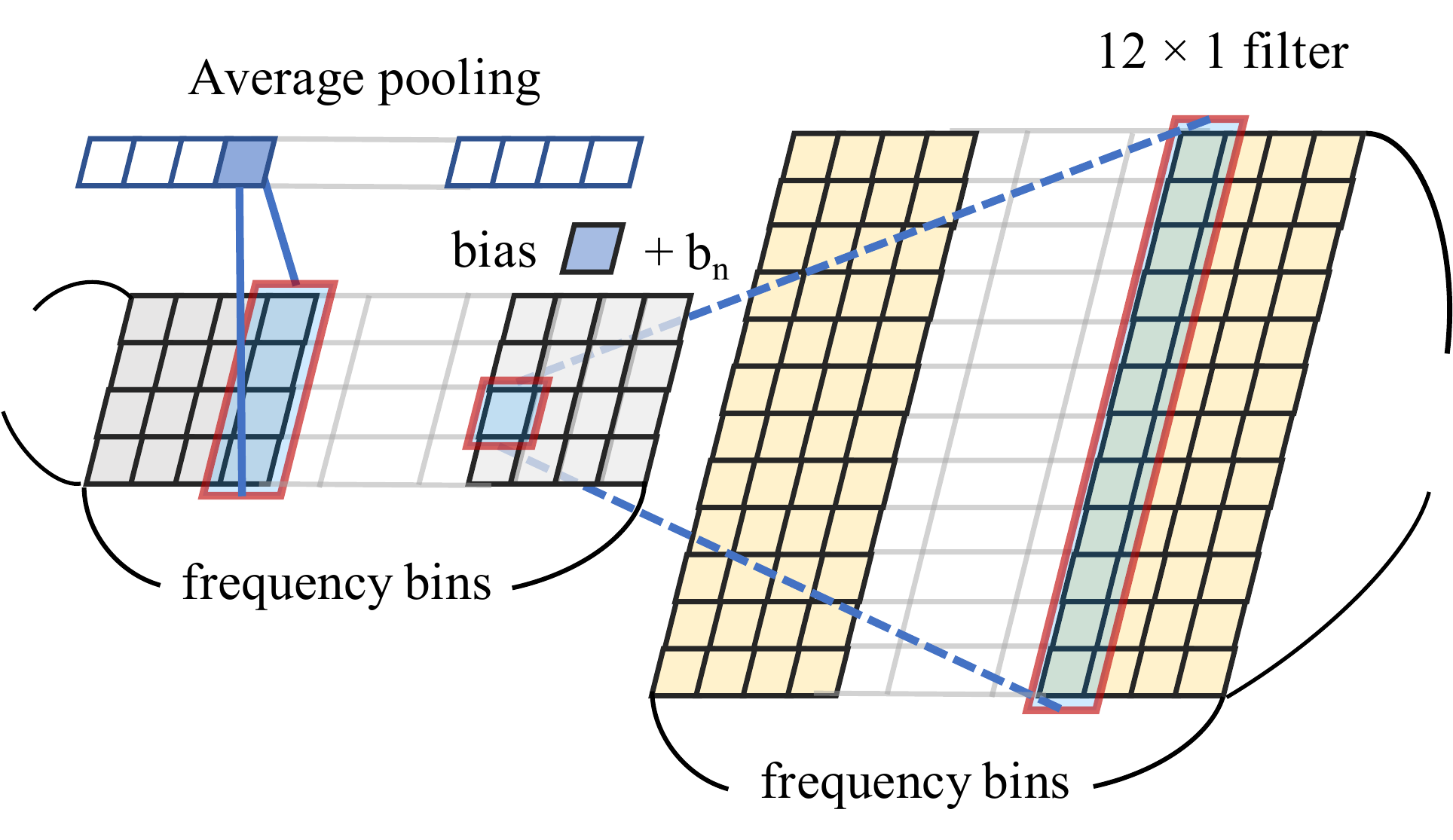}}
    \centerline{}
    \vspace{-4.0ex}
\caption{Example of FAN architecture for 12 dimensional input and 4 filters.}
    \vspace{-3.0ex}
\label{fig:FAN}
\end{figure}



\subsection{Frequency Aligned Network}
BAT processes frequency bins independently as it implements filtering in the frequency domain. Applying an affine transform after BAT can impact the overall performance as a typical affine transform (linear layer) combines the  outputs of various frequency bins. Note that this combination of frequency bins is different from the output of the feature extraction network in the single channel case depicted in Fig \ref{fig:baseline}. The feature extraction scheme here simply implements filter-band energy calculation. In this work,  we seek a new solution to the frequency combination problem for the multi-channel spatial filtering built into the network. Ideally, we seek an approach that handles each frequency bin independently across various look directions.

The fundamental idea of FAN proposed here is all about processing each frequency bin separately. Fig.\ref{fig:FAN} depicts an example of the FAN with 12 look directional input from BAT and 4 filters. In Fig.\ref{fig:FAN}, each filter has 12 $\times$ 1 coefficient that is multiplied by each frequency. The 12 coefficients in each filter can be regarded as a weight for each look direction. The output of the filtering process for each frequency bin is handled separately.

The overall filtering structure of FAN is analogous to convolutional neural network (CNN) but FAN architecture does not involve explicit convolution. A 1-D CNN  layer typically involves convolution by shifting the filter with a certain hop length; each output is affected by neighboring inputs as convolution process involves overlap and shifting. In contrast, in the FAN architecture, there is no overlap when the filter is shifted to the neighboring frequency bin. FAN is designed to effectively weigh look directions for each frequency bin.

FAN should be also distinguished from BAT. BAT returns multiple of look directions for varoious frequency bins. Specifically, BAT combines phase information to provide multiple angles of acoustic spatial filtering, while FAN is only used to weigh each look direction and combine. The output from eq.\ref{eq:bat} forms an input vector for frequency with $\omega_{k}$ in eq.\ref{eq:yk}.

\begin{equation}
\label{eq:yk}
\mathbf{Y}\left(\omega_{k}\right)=\left[ Y_{1}\left(\omega_{k}\right), Y_{2}\left(\omega_{k}\right), \dots, Y_{D}\left(\omega_{K}  \right)\right]
\end{equation} 
Using the input vector $\mathbf{Y}\left(\omega_{k}\right)$ the output of FAN can be described as follows: 

\begin{equation}
Z(\omega_k)=\frac{1}{N} \sum_{n=1}^{N} \left( \mathbf{w}_{FAN, n}^{H} \mathbf{Y}(\omega_{k}) + b_{n} \right)
\end{equation}
where $\mathbf{w}_{FAN, n}$ is n-th filter coefficient, $b_n$ is bias term for n-th filter and $N$ is the total number of filters where we use 24 filters. The output $Z(\omega_k)$ directly goes into the FE-DNN in Fig.\ref{fig:baseline}. In the case of FAN-Max model in Fig.\ref{fig:baseline}, the dimension of input to FAN is D=2 since we only use 2-ch input instead of 12 directional looks. In the case of BAT-FAN-Max model, average pooling is replaced with max pooling to gauge the benefit of the pooling method. 

 In addition to frequency independent processing for combining look directions, FAN requires significantly fewer parameters in comparison to an affine transform. In our case,  an affine transform requires 193,548 (12 $\times$ 127 $\times$ 127 + 127) parameters while FAN architecture only requires 192 (12 $\times$ 24 + 24) parameters. 

\subsection{MC modules}
\begin{figure}[t]
  \centering
  \centerline{\includegraphics[width=8.3cm]{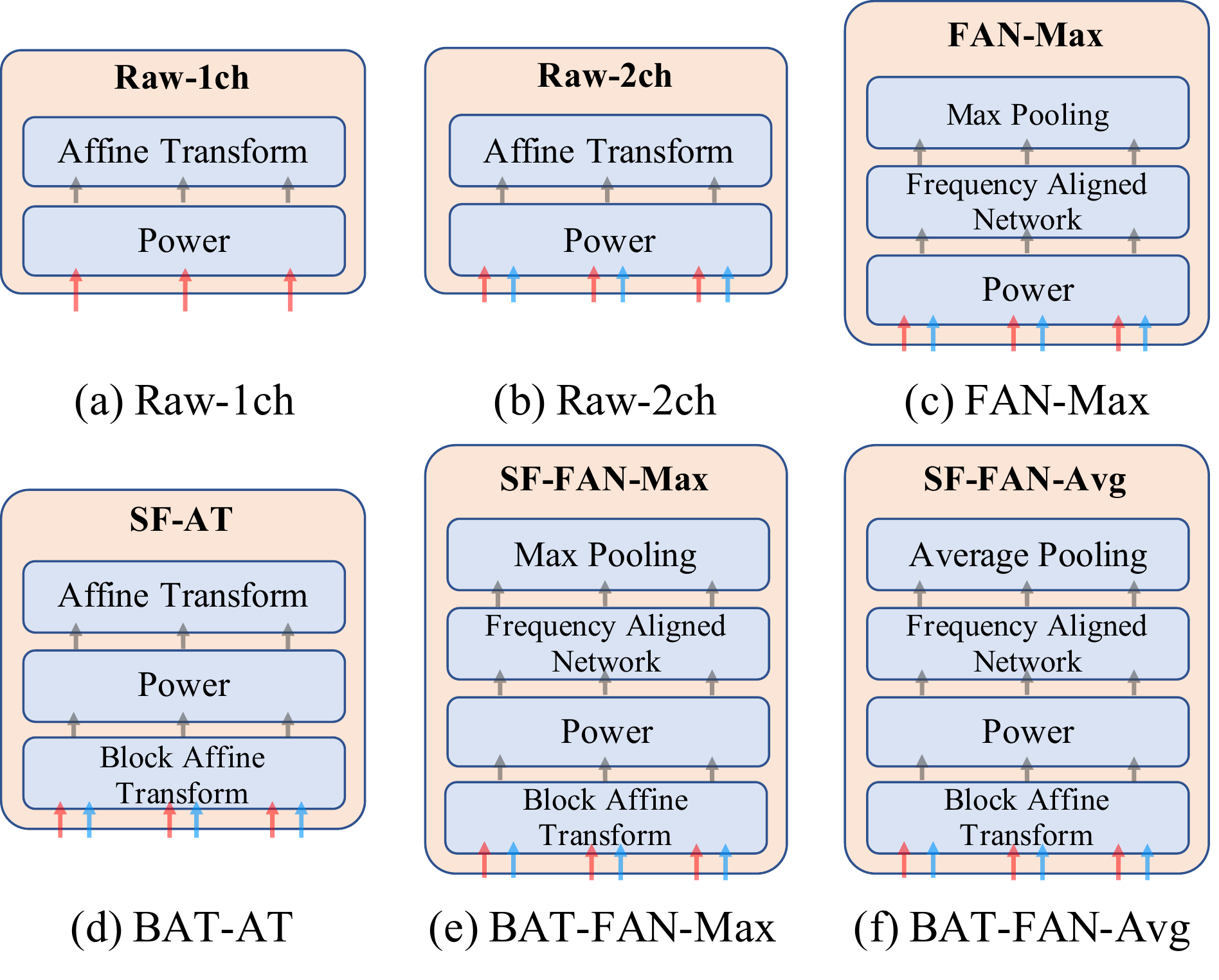}}
    \centerline{}
    \vspace{-3.0ex}
\caption{MC modules with different sub-components.}
\label{fig:mc_modules}
    \vspace{-3.0ex}
\end{figure}

MC module in our DSR framework (Fig. \ref{fig:baseline}) is core to the experiments presented here.  Fig.\ref{fig:mc_modules}. describes the 6 types of MC modules we investigate. The modules in Fig.\ref{fig:mc_modules} can be plugged into the baseline model shown in Fig. \ref{fig:baseline} and jointly trained.

The first row of Fig.\ref{fig:mc_modules} shows the MC architecture without BAT and therefore do not have spatial filtering capability. Fig.\ref{fig:mc_modules}(a) `Raw 1-ch' is a model that only uses single channel input. This model is the pretrained network we use to train the classification network; except that here in Fig.\ref{fig:mc_modules}(a), we use the FE-DNN component for feature extraction for the classification network. The power component computes and the pair-wise sum of squares that halve the dimension. In this work, 254 dimensional DFT becomes 127 dimensional output after the power operation. The affine transform in Fig.\ref{fig:mc_modules}(a)  then converts 127 dimensional input to the same 127 dimensional output.  The 2-channel model in Fig.\ref{fig:mc_modules}(b) `Raw-2ch' depicts a 2-channel version of the  Raw 1-ch model. This model is primarily designed to test the benefit of BAT and FAN. The affine transform in Raw-2ch has 254 dimensional input and 127 dimensional output. The `FAN-Max' model in Fig.\ref{fig:mc_modules}(c) is a model that comprises FAN and Max pooling module instead of affine transform in Raw-2ch model. FAN-Max model is designed to examine the benefit of having BAT module in our DSR framework. In FAN-Max model, FAN architecture acceptes 2 by 127 input and has filter dimension of 2 by 1.

The second row of Fig.\ref{fig:mc_modules} shows MC modules with spatial filtering capability (with BAT). The `BAT-AT' model in Fig.\ref{fig:mc_modules}(d) is the best performing system from \cite{minhua2019frequency} which is referred as \emph{elastic special filtering}. The `BAT-FAN-Max' model in Fig.\ref{fig:mc_modules}(e) is a model which we integrate BAT with FAN architecture. The affine transform in BAT-AT model is replaced with FAN and max pooling module. Finally, `BAT-FAN-Avg' model in Fig.\ref{fig:mc_modules}(f) is a model with average pooling instead of max pooling. We can see the benefit from having a average pooling layer by comparing the performance of BAT-AT model. BAT module and FAN architecture is explained in more detail in the following sections.

\begin{table*}[t]
\scriptsize
\caption{Word Error Rate Reduction relative to  Raw-1ch model by SNR level}
\vspace{2ex}
\label{tab:table1}
\centering
\begin{tabular}{ c c || c c c c || c c c c || c c c c }
\hline 
&  
& \multicolumn{4}{c||}{\textbf{SNR $\leq$ 5dB }}
& \multicolumn{4}{c||}{\textbf{5dB $<$ SNR $\leq$ 15dB }}
& \multicolumn{4}{c }{\textbf{15dB $<$ SNR}} \\

Systems & Split 
&  Set1 noPB  &  Set1 PB & Set2 PB & Total
&  Set1 noPB  &  Set1 PB & Set2 PB & Total
&  Set1 noPB  &  Set1 PB & Set2 PB & Total \\

\hline 

\multirow{2}{*}{\textbf{Raw-2ch}} & dev & 8.48&  10.89&  6.18&  6.68&  7.63&  8.70&  7.34&  7.14&  5.26&  13.25&  0.00&  4.39 \\
									   & test &  3.88&  7.00&  5.25&  5.18&  7.34&  8.66&  5.06&  6.40&  4.63&  7.48&  5.93&  4.59 \\ \hline
\multirow{2}{*}{\textbf{FAN-Max}} & dev & 6.70&  10.12&  7.13&  7.22&  6.87&  0.00&  7.34&  6.49&  4.39&  3.61&  5.71&  5.26 \\
 									  & test & 2.43&  11.11&  6.13&  5.79&  5.50&  5.51&  7.30&  5.60&  3.70&  1.87&  6.78&  3.67 \\ \hline
\multirow{2}{*}{\textbf{BAT-AT}}& dev & 10.27&  12.06&  6.65&  7.22&  10.69&  20.50&  9.60&  10.39&  9.65&  10.84&  4.76&  8.77 \\
 									  & test &  5.83&  14.40&  7.44&  7.93&  9.17&  7.87&  7.30&  8.00&  9.26&  6.54&  8.47&  8.26 \\ \hline
\multirow{2}{*}{\textbf{BAT-FAN-Max}}& dev & 14.29&  17.51&  13.78&  14.17&  17.56&  18.63&  15.25&  16.23&  15.79&  21.69&  17.14&  16.67 \\
									  & test & 9.22&  19.75&  15.10&  14.02&  14.68&  22.05&  16.29&  15.20&  15.74&  8.41&  15.25&  14.68 \\ \hline
\multirow{2}{*}{\textbf{BAT-FAN-Avg}}& dev &  17.41&  25.68&  18.29&  18.72&  19.85&  26.71&  20.90&  20.78&  20.18&  22.29&  19.05&  20.18 \\
									  & test &  13.59&  25.93&  17.29&  17.38&  19.27&  26.77&  19.66&  20.00&  19.44&  14.95&  21.19&  18.35 \\ \hline

\hline 


\end{tabular}
\vspace{-3.0ex}
\normalsize
\end{table*}
\vspace{-2.0ex}
\section{EXPERIMENTAL RESULTS}

\begin{table}
\scriptsize
\caption{Word Error Rate Reduction relative to  Raw-1ch model}
\vspace{2ex}
\label{tab:table2}
\centering
\begin{tabular}{ c c ||  c | c | c  | c }
\hline 
&  
& \textbf{Set1 noPB }
& \textbf{Set1 PB }
& \textbf{Set2 PB }
& \textbf{Total} \\

\hline
\multirow{2}{*}{\textbf{Raw-2ch}} & dev & 6.92 & 10.71 & 	5.98 & 	\textbf{6.36}  \\
 									  & test &  5.29 & 	7.45 & 	5.28 & 	\textbf{5.37}  \\ \hline
\multirow{2}{*}{\textbf{FAN-Max}} & dev & 5.83 & 	6.33 & 	7.07 & 	\textbf{6.66}  \\
 										& test & 3.87 & 	8.48 & 	6.42 & 	\textbf{5.25} \\ \hline
\multirow{2}{*}{\textbf{BAT-AT}} & dev & 10.12 & 14.12 & 7.11 & \textbf{8.45} \\
 								 & test & 8.14 & 11.75 & 7.52 & 	\textbf{7.99}  \\ \hline
\multirow{2}{*}{\textbf{BAT-FAN-Max}} & dev & 16.02 & 	18.50 & 14.33 & \textbf{15.25} \\
								 & test & 13.35 & 	18.74 & 15.35 & \textbf{14.47} \\ \hline
\multirow{2}{*}{\textbf{BAT-FAN-Avg}} & dev & 19.30 & 25.41 & 18.85 & \textbf{19.57} \\
 								& test & 17.61 & 24.58 & 18.11 & \textbf{18.32}  \\ \hline

\hline 


\end{tabular}
\vspace{-3.0ex}
\normalsize
\end{table}
\label{sec:pagestyle}
\subsection{Experiment setup}
We examine the performance of the proposed MC modules with multi-channel speech dataset. The dataset was collected from several thousand anonymized users using 7 microphone circular array devices in real acoustic environments. We use two types of datasets. Set1 is a dataset which contains utterances recorded in real world conditions with playback echo for approximately 10\% of the whole data. In this case, the volunteers involved in the data collection could set the playback volume to an arbitrary values. Set2 is an in-lab dataset where all the utterances in  Set2 were always recorded with playback echo.  No acoustic echo cancellation (AEC) was applied. 
The dev set and test set have approximately 20 hours and 50 hours of data. 
 The array geometry used here is an equi-spaced six-channel microphone circular array with a diameter of approximately 72 mm and one microphone at the center. All the experiments in this work are  two channel experiments where we picked two microphones diagonally across the center of the circular array \cite{minhua2019frequency}. The speech data in the training set and test set are device-directed speech data where the user's speech is directed toward the device.  

For all the experiments, we use DFT feature. We remove the direct and Nyquist frequency components in the DFT coefficient and use 127-dimensional DFT coefficients. The DFT feature was extracted every 10ms using a window size of 12.5 ms (and padded). We normalize DFT feature with the global mean and variances that are precomputed from the training data. The LSTM in classification network has 5 LSTM layers with 768 cells followed by the affine transform that has 3183 dimensional output. We use the cross-entropy loss using a neural network toolkit \cite{strom2015scalable}. Adam optimizer is used in all the experiments. 

To illustrate the benefit of the various modules outlined earlier, we run ASR experiments on the dataset we collected and the results are presented in terms of word error rate reduction (WERR). As we take an additive intereference view of the playback signal and no AEC was applied, the WERR results are broken down into estimated signal-to-noise ratio (SNR) of the utterances in dev set and test set. The SNR is estimated by aligning the utterances to the transcriptions using an offline ASR model \footnote{This offline model was trained on beamformed audio with AEC} and calculating the ratio between accumulated power of speech and noise frames from a whole utterance. 

\subsection{Performance comparison}

Table.\ref{tab:table1} shows WERRs  of the different models evalated in this work. The hyperparameters (e.g. number of filters for FAN) were optimized on dev set and finally evaluated on test split. `noPB' indicates the subset of dataset where playback echo was not present and `PB' represents the subset of  utterances recorded in the presence of playback echo. Performance on Set2 and combined Set1+2 PB total (for utterances with playback echo) is also shown.

From the table, it can be seen that Raw-2ch model shows consistent improvement over Raw-1ch model. The Raw-2ch model's WERR indicates that by  adding an additional input channel can improve ASR performance. In addition, Raw-2ch model's result can be compared with FAN-Max model, where the only difference is FAN and affine transform. Note that FAN-Max model does not have additional benefit from DFT feature since DFT signal directly goes through power operation and then fed into FAN. However, we can see that FAN-Max model performs better in low SNR while Raw-2ch model performs better in higher SNR conditions. 

The BAT-AT model is a more competitive baseline that shows consistent benefit over Raw-2ch model since BAT-AT model uses spatial filtering prior knowledge \cite{minhua2019frequency}. In comparison, the performances of BAT-FAN-Max model and BAT-FAN-Avg model show the benefit of FAN architecture.  BAT-FAN-Max model shows  about 7\% WERR improvement over the BAT-AT model, which indicates that there is a benefit of  frequency independent processing.  There is also benefit of average pooling method with FAN compared to the model that uses maxpool for both dev and test sets. Note that BAT-FAN-Avg model shows significant improvements especially for low SNR conditions. 

Table \ref{tab:table2} shows overall WERR of the MC modules. A noteworthy result in Table \ref{tab:table2} is the improvement gap between Set1 noPB and Set1 PB. We can see that the improvement from FAN and average pooling is on Set1 PB, while Set2 PB has slightly better WERR over Set1 noPB. This contrasts to BAT-AT model's improvement in Set1 PB and Set2 PB since the improvements are less concentrated on the utterances with playback echo. According to this experimental results, the proposed FAN architecture followed by average pooling gives even more benefit to the utterances with intense noise.

\section{Conclusion}
In this paper, we presented  frequency-aligned architecture for multi-channel acoustic modeling. The experimental results suggest that frequency independent processing largely improves the performance of our distant speech recognition system. In addition, this architecutre has many orders fewer parameters than linear layer while showing improved performance. We also found that the subsequent step of average- versus max-pooling  can result in  further improvement on top of the the frequency aligned architecture.  In addition to evaluating this architecture on a larger dataset, our future work  would include cascaded layers of frequency-aligned networks and  other configuraitions for a 4-channel  and multi-geometry \cite{Kumatani2019multigeometry} acoustic model.
\vfill\pagebreak
\bibliographystyle{IEEEbib}
\bibliography{refs}

\begin{thebibliography}{10}

\bibitem{delcroix2014linear}
Marc Delcroix, Takuya Yoshioka, Atsunori Ogawa, Yotaro Kubo, Masakiyo Fujimoto,
  Nobutaka Ito, Keisuke Kinoshita, Miquel Espi, Takaaki Hori, Tomohiro
  Nakatani, et~al.,
\newblock ``Linear prediction-based dereverberation with advanced speech
  enhancement and recognition technologies for the reverb challenge,''
\newblock in {\em Reverb workshop}, 2014.

\bibitem{kumatani2012microphone}
Kenichi Kumatani, John McDonough, and Bhiksha Raj,
\newblock ``Microphone array processing for distant speech recognition: From
  close-talking microphones to far-field sensors,''
\newblock {\em IEEE Signal Processing Magazine}, vol. 29, no. 6, pp. 127--140,
  2012.

\bibitem{sullivan1996multi}
Thomas~M Sullivan,
\newblock ``Multi-microphone correlation-based processing for robust automatic
  speech recognition,''
\newblock {\em Unpublished doctoral dissertation, Carnegie Mellon University,
  Pittsburgh, USA}, 1996.

\bibitem{himawan2010clustered}
Ivan Himawan, Iain McCowan, and Sridha Sridharan,
\newblock ``Clustered blind beamforming from ad-hoc microphone arrays,''
\newblock {\em IEEE Transactions on Audio, Speech, and Language Processing},
  vol. 19, no. 4, pp. 661--676, 2010.

\bibitem{mcdonough2008distant}
John McDonough and Matthias Wolfel,
\newblock ``Distant speech recognition: Bridging the gaps,''
\newblock in {\em 2008 Hands-Free Speech Communication and Microphone Arrays}.
  IEEE, 2008, pp. 108--114.

\bibitem{seltzer2008bridging}
Michael~L Seltzer,
\newblock ``Bridging the gap: Towards a unified framework for hands-free speech
  recognition using microphone arrays,''
\newblock in {\em 2008 Hands-Free Speech Communication and Microphone Arrays}.
  IEEE, 2008, pp. 104--107.

\bibitem{xiao2016deep}
Xiong Xiao, Shinji Watanabe, Hakan Erdogan, Liang Lu, John Hershey, Michael~L
  Seltzer, Guoguo Chen, Yu~Zhang, Michael Mandel, and Dong Yu,
\newblock ``Deep beamforming networks for multi-channel speech recognition,''
\newblock in {\em 2016 IEEE International Conference on Acoustics, Speech and
  Signal Processing (ICASSP)}. IEEE, 2016, pp. 5745--5749.

\bibitem{ochiai2017multichannel}
Tsubasa Ochiai, Shinji Watanabe, Takaaki Hori, and John~R Hershey,
\newblock ``Multichannel end-to-end speech recognition,''
\newblock in {\em Proceedings of the 34th International Conference on Machine
  Learning-Volume 70}. JMLR. org, 2017, pp. 2632--2641.

\bibitem{swietojanski2014convolutional}
Pawel Swietojanski, Arnab Ghoshal, and Steve Renals,
\newblock ``Convolutional neural networks for distant speech recognition,''
\newblock {\em IEEE Signal Processing Letters}, vol. 21, no. 9, pp. 1120--1124,
  2014.

\bibitem{braun2018multi}
Stefan Braun, Daniel Neil, Jithendar Anumula, Enea Ceolini, and Shih-Chii Liu,
\newblock ``Multi-channel attention for end-to-end speech recognition.,''
\newblock in {\em Interspeech}, 2018, vol. 2018, pp. 17--21.

\bibitem{kim2015recurrent}
Suyoun Kim and Ian Lane,
\newblock ``Recurrent models for auditory attention in multi-microphone
  distance speech recognition,''
\newblock {\em arXiv preprint arXiv:1511.06407}, 2015.

\bibitem{heymann2018performance}
Jahn Heymann, Michiel Bacchiani, and Tara~N Sainath,
\newblock ``Performance of mask based statistical beamforming in a smart home
  scenario,''
\newblock in {\em 2018 IEEE International Conference on Acoustics, Speech and
  Signal Processing (ICASSP)}. IEEE, 2018, pp. 6722--6726.

\bibitem{higuchi2018frame}
Takuya Higuchi, Keisuke Kinoshita, Nobutaka Ito, Shigeki Karita, and Tomohiro
  Nakatani,
\newblock ``Frame-by-frame closed-form update for mask-based adaptive mvdr
  beamforming,''
\newblock in {\em 2018 IEEE International Conference on Acoustics, Speech and
  Signal Processing (ICASSP)}. IEEE, 2018, pp. 531--535.

\bibitem{minhua2019frequency}
Wu~Minhua, Kenichi Kumatani, Shiva Sundaram, Nikko Str{\"o}m, and Bj{\"o}rn
  Hoffmeister,
\newblock ``Frequency domain multi-channel acoustic modeling for distant speech
  recognition,''
\newblock in {\em ICASSP 2019-2019 IEEE International Conference on Acoustics,
  Speech and Signal Processing (ICASSP)}. IEEE, 2019, pp. 6640--6644.

\bibitem{pundak2016lower}
Golan Pundak and Tara Sainath,
\newblock ``Lower frame rate neural network acoustic models,''
\newblock 2016.

\bibitem{kumatani2017direct}
Kenichi Kumatani, Sankaran Panchapagesan, Minhua Wu, Minjae Kim, Nikko Strom,
  Gautam Tiwari, and Arindam Mandai,
\newblock ``Direct modeling of raw audio with dnns for wake word detection,''
\newblock in {\em 2017 IEEE Automatic Speech Recognition and Understanding
  Workshop (ASRU)}. IEEE, 2017, pp. 252--257.

\bibitem{richard2013perceptual}
G.~{Richard}, S.~{Sundaram}, and S.~{Narayanan},
\newblock ``An overview on perceptually motivated audio indexing and
  classification,''
\newblock {\em Proceedings of the IEEE}, vol. 101, no. 9, pp. 1939--1954, Sep.
  2013.

\bibitem{strom2015scalable}
Nikko Strom,
\newblock ``Scalable distributed dnn training using commodity gpu cloud
  computing,''
\newblock in {\em Sixteenth Annual Conference of the International Speech
  Communication Association}, 2015.

\bibitem{Kumatani2019multigeometry}
K.~{Kumatani}, W.~{Minhua}, S.~{Sundaram}, N.~{Ström}, and B.~{Hoffmeister},
\newblock ``Multi-geometry spatial acoustic modeling for distant speech
  recognition,''
\newblock in {\em ICASSP 2019 - 2019 IEEE International Conference on
  Acoustics, Speech and Signal Processing (ICASSP)}, May 2019, pp. 6635--6639.

\end{thebibliography}

\end{document}